\documentclass[12pt]{article}

\usepackage{amsmath, amssymb, graphicx, psfrag}
\usepackage{caption}
\usepackage{float}
\usepackage{subcaption}
\usepackage{physics, mathtools, subcaption, empheq}
\usepackage[numbers,authoryear,round]{natbib}
\usepackage{titlesec}
\usepackage{hyperref}
\hypersetup{breaklinks=true}
\bibpunct{(}{)}{;}{a}{,}{,}
\makeatletter
\renewcommand{\NAT@sep}{;\penalty0\ }   
\renewcommand{\NAT@aysep}{,\penalty0\ } 
\makeatother
\hypersetup{
    colorlinks = true,
    urlcolor   = blue,
    citecolor  = black,
}

\title{Whistling of turbulent cavity flows: Self-consistent predictions with the linearized compressible Navier-Stokes equations}
\author{Nikolaos Bozikis,  Dilara Özev and Nicolas Noiray
}
\date{CAPS Laboratory, Department of Mechanical and Process Engineering, ETH Zürich, Zürich 8092,
Switzerland}

\begin{document}
\maketitle

\begin{abstract}
A self-consistent saturation model for the prediction of aeroacoustic limit cycles emerging in turbulent low-Mach cavity flows ($\text{Re}=\mathcal{O}(10^5)$, $M\simeq0.2$) is proposed. It predicts the nonlinear interactions between the acoustic modes of a deep rectangular cavity and the hydrodynamic instabilities of the turbulent shear-layer that forms over its opening due to the presence of a grazing flow. The model is based on the triple decomposition of the flow variables and the compressible Navier-Stokes equations. At each step of the iterative process, the nonlinear eigenvalue problem associated to perturbations around the mean flow is updated with the steady component of the forcing from the unstable eigenmode's Reynolds stress. The iterations are performed until the dominant eigenmode becomes marginally stable, i.e. its growth rate vanishes. The evolution of the coherent velocity fluctuations as function of the oscillation amplitude is in good qualitative agreement with previously published compressible Large Eddy Simulations. Furthermore, the predictions of the frequency and amplitude of the aeroacoustic limit cycle oscillations are validated against the ones obtained from a low order model, whose parameters were adjusted to reproduce the experimental measurements of the deep cavity whistling.
\end{abstract}

\section{Introduction}
\label{intro}
    Low-Mach grazing flows over cavities can lead to constructive interactions between hydrodynamic modes of the shear layer and acoustic modes of the cavity. These interactions can result in high amplitude aeroacoustic limit cycle oscillations, which potentially induce cyclic stresses leading to structural fatigue. The prediction of the occurrence of such whistling phenomenon and its intensity are therefore highly relevant for practical applications such as aeronautical engineering, ground-based transportation systems, and turbo-machinery. The shear layer at the cavity opening is prone to Kelvin-Helmholtz instabilities, leading to the amplification of vorticity oscillations originating from the upstream corner of the opening. Their interaction with the acoustic modes of the cavity creates a positive feedback loop that leads to self-sustained aeroacoustic oscillations, e.g. \citet{article}, \citet{Bourquard_Faure-Beaulieu_Noiray_2021}, \citet{Ho_Kim_2021} and \citet{Wang_Jia_He_He_Sung_Liu_2024}. In the case of turbulent grazing flow, \citet{Bourquard_Faure-Beaulieu_Noiray_2021} showed that stochastic perturbation from turbulence can lead to intermittent instability and derived the coupled Langevin equations governing the aeroacoustic system in such situation. Recent experiments \citep{HANNA2023103949,Faure-Beaulieu_Xiong_Pedergnana_Noiray_2023} and Large Eddy Simulations \citep{saturation_boujo_2018,Ho_Kim_2021,10.1063/5.0051226}, uncovered new aspects of governing mechanisms leading to these so-called fluid-resonant oscillations \citep{Rockwell}. However the latter simulations require large computational costs, which hinders their use for multiple-step engineering design optimization. Consequently, there is a need for reduced order models such as the Discrete Vortex Model (DVM) implemented by \citet{article} with a modified version of the Rossiter mechanism \citeyearpar{rossiter1964wind}, or   decoupled Computational Aero-Acoustics (CAA) methods, e.g.  \citet{koh2003aeroacoustic}.  Low-order models with experimentally determined constants have also been developed such as the empirical model of \citet{Bourquard_Faure-Beaulieu_Noiray_2021} that is able to predict whistling frequencies and limit cycle amplitudes for the deep cavity subjected to low-Mach turbulent grazing flow system.

 Linear stability analysis around a mean flow, while unable to capture the finite-amplitude interactions leading to the development of limit-cycle oscillations, can provide valuable insights on the system's stability characteristics. For instance,  \citet{Yamouni_Sipp_Jacquin_2013} performed a global stability analysis of the compressible Navier-Stokes equations and were able to identify two destabilizing mechanisms of the flow over a cavity. \citet{Citro_Giannetti_Brandt_Luchini_2015} used asymptotic stability analysis, assuming the short-wavelength approximation, to study and categorize the mechanisms of instability of incompressible cavity flow. \citet{Sierra-Ausin_Fabre_Citro_Giannetti_2022} considered the flow of a jet through a circular cavity and constructed an matched asymptotic model to predict acoustic instabilities via an impedance criterion. The impedances necessary for closure were derived from the linearized Navier-Stokes equations (LNSE). In the work of \citet{saturation_boujo_2018}, the incompressible Navier-Stokes equations are linearized around the mean flow over a deep cavity subjected to external forcing in order to perform a sensitivity analysis of thew flow and identify the mechanisms governing its coherent response. In particular, a saturation phenomenon is observed, where increasing the forcing amplitude reduces the amplitude of the system's response (harmonic gain) to the forcing.

  In the present study a self-consistent analysis of the same flow configuration as the one of \citet{saturation_boujo_2018} and \citet{Bourquard_Faure-Beaulieu_Noiray_2021} is performed to describe the emergence of finite amplitude aeroacoustic limit cycles. The saturation amplitude and frequency predictions of the model are compared to experimental data and the accuracy of the method is assessed. The self-consistent approach used here is an extension of the work of \citet{PhysRevLett.113.084501} to the case of a low-Mach, turbulent and compressible flow. \citet{Stuart_1958} was the first to propose an analysis of the linear stability of a mean flow for different amplitudes of a linearly unstable mode. More recently, \citet{PhysRevLett.113.084501} proposed to solve the mean flow equation iteratively with the updated steady forcing from the unstable mode at each iteration. The idea of iteratively updating the steady forcing field was a crucial improvement to the method proposed by \citet{Stuart_1958} because the eigenmode shape usually strongly depends on the mean flow. Eventually the mean flow becomes marginally stable rendering the growth rate of the instability zero. This indicates a balance between linear and nonlinear effects and the system is then considered to be saturated, e.g. \citet{Meliga_2017}.
  \citet{PhysRevLett.113.084501} demonstrated that their self-consistent model can accurately predict the amplitude and frequency of limit cycle oscillations, in the canonical case of periodic vortex shedding in the laminar wake behind a cylinder without reliance on DNS or experimental data for calibration. \citet{mantic2016self} adapted the self-consistent model to analyze the saturation of the response to harmonic forcing in a laminar backward-facing step flow. In the work of \citet{AcousticCyl} the full compressible NS equations were considered to investigate and predict the acoustic far field caused by the hydrodynamic modes of a laminar oscillating wake behind a cylinder. \citet{yim2019self} performed a triple decomposition of the flow and supplemented the incompressible NS with an eddy viscosity model thus accounting for the effects of turbulence. The success of self-consistent saturation models is not limited to the accurate prediction of the saturated base flow field and the limit cycle's amplitude and frequency. It also has the ability to highlight the underlying physical mechanisms leading to saturation i.e. the modifications imposed on the base flow by the developing mode, from the onset of instability up to saturation.
  In \citet{Li_Yang_2025} the implementation of self-consistent models in the development of active wake control (AWC) strategies is proposed, after their potential to significantly reduce the computational costs of wake calculations is demonstrated in their work.  
  
  In the present work an extended version of the self-consistent model is proposed. It accounts for low-Mach compressibility effects as well as the effects of high Reynolds turbulence, enabling predictions of aero-acoustic whistling phenomena. The problem is presented in the next section alongside a full description of the methods used. In the last section the results are presented and validated against experiments and compressible Large Eddy Simulations.

\section{Problem description}
\subsection{Flow Domain}
\begin{figure}
    \centering
    \psfrag{H}[][][0.8]{$H$}
    \psfrag{W}[][][0.7]{$W$}
    \psfrag{0}[][][0.6]{$0$}
    \psfrag{7}[][][0.6]{$70$}
    \psfrag{ms}[][][0.6]{(m s$^{-1}$)}
    \psfrag{Ub}[][][0.7]{$\overline{u}$}
    \psfrag{ls}[][][0.8]{$l_s$}
    \psfrag{x}[][][0.7]{$\hat{x}$}
    \psfrag{Y}[][][0.7]{$\hat{y}$}
    \psfrag{L1}[][][0.8]{$L_1$}
    \psfrag{L2}[][][0.8]{$L_2$}
    \psfrag{Gi}[][][0.8]{\color[RGB]{8, 28, 194}{$\Gamma_\text{in}$}}
    \psfrag{Go}[][][0.8]{\color[RGB]{221, 22, 22}{$\Gamma_\text{out}$}}
    \psfrag{Gw}[][][0.8]{$\Gamma_\text{wall}$}
    \psfrag{Gs}[][rd][0.8]{\color[RGB]{43, 120, 9}{$\Gamma_\text{slip}$}}
    \psfrag{D}[][][0.8]{$D$}
    \psfrag{U}[][][0.8]{\color[RGB]{8, 28, 194}{$U_{\infty}$}}
    \includegraphics[width=0.7\linewidth]{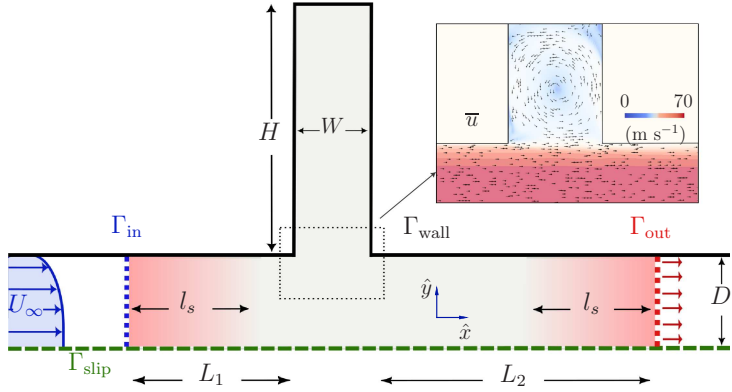}
    \caption{Flow domain sketch. The acoustically dampening buffer zones of thickness $l_s$ are indicated by red color. Inset: Contour of mean velocity magnitude $\overline{u}$ with overlayed arrows indicating direction.}
    \label{Domain}
\end{figure}

 The 2D flow domain consists of a straight duct that features a closed cavity side branch of rectangular shape as shown in figure \ref{Domain}. The duct cross section length is $D=6.2$ cm, while the cavity measures $W=3$ cm in width and its height $H$ is varied in this work between $18$ and $30$ cm. Air  flows inside the duct with a bulk velocity $U_b$ varied between 55 to $70\ \text{m\,s}^{-1}$. This choice of parameters yields a set of flows characterized by small Mach and large Reynolds numbers, $U_b/c_0=M\in [0.16,\,0.2]$ where $c_0=340\ \text{m\,s}^{-1}$ is the speed of sound and $U_b W/\nu=\text{Re} \simeq 1.5\times10^{5}$ where $\nu=1.5\times10^{-5} \ \text{m}^2\,\text{s}^{-1}$ is the kinematic viscosity, while the Prandtl number $\text{Pr}=\rho_0 \nu c_p/ \kappa$ is around 0.7, where $\rho_0=1.2\,\text{kg m}^{-3}$ is the density, $c_p=10^3 \ \text{J\,kg}^{-1}\ \text{K}^{-1}$ the specific heat capacity and $\kappa=2.6\times10^{-2} \ \text{J\,s}^{-1}\text{m}^{-1}\text{K}^{-1}$ the thermal conductivity.

\subsection{Governing equations}
 The compressible Navier-Stokes equations are cast in non-dimensional form using the reference values presented above, as in the work of \citet{AcousticCyl} and \citet{MELIGA_SIPP_CHOMAZ_2010} on noise prediction from flow over a bluff body and a shallow cavity respectively:
\begin{equation}
    \partial_t \varrho + \grad\cdot(\varrho\boldsymbol{u})=0,
\end{equation}
\begin{equation}
    \varrho\partial_t \boldsymbol{u} + \boldsymbol{N}(\varrho,\boldsymbol{u},P)=0,
\end{equation}
$$ \boldsymbol{N}(\varrho,\boldsymbol{u},P)=\varrho(\boldsymbol{u}\cdot\grad)\boldsymbol{u} + \frac{1}{\gamma M^2} \grad P - \frac{1}{\text{Re}}\grad\cdot\boldsymbol{\tau}(\boldsymbol{u})$$
where $\gamma$ is the ratio of specific heat capacities. Furthermore, the non-dimensionalized energy equation reads
\begin{equation}
    \varrho\partial_tT + \boldsymbol{E}(\varrho,\boldsymbol{u}, T)=0,
\end{equation} with
$$ \boldsymbol{E}(\varrho,\boldsymbol{u}, T) = \varrho(\boldsymbol{u}\cdot\grad)T + (\gamma-1)\varrho T\grad\cdot\boldsymbol{u} - \gamma(\gamma-1)\frac{M^2}{\text{Re}}\boldsymbol{\tau}(\boldsymbol{u}):\boldsymbol{d}(\boldsymbol{u}) -\frac{\gamma}{\text{Pr}\text{Re}}\grad^2T .$$

Closure is achieved by means of the equation of state $P = \varrho T$, 
where the non-dimensional pressure  $P$ is defined as $(p-p_0)/(\rho_{0}U_b^2)$. The solution $\boldsymbol{q}=[\varrho,\boldsymbol{u},T,P]^{\intercal}$ is decomposed into its time averaged component $\overline{\boldsymbol{q}}(\boldsymbol{x})$, coherent fluctuations $\widetilde{\boldsymbol{q}}(\boldsymbol{x},t)$ and turbulent fluctuations $\boldsymbol{q}'(\boldsymbol{x},t)$ following \citet{Hussain_Reynolds_1970}, where the different components can be isolated from the complete unsteady flow by phase averaging $(\langle \cdot\rangle)$ and time averaging $(\overline{\cdot})$
\begin{equation}
    \boldsymbol{q}=\overline{\boldsymbol{q}}+\widetilde{\boldsymbol{q}}+\boldsymbol{q}'
    \label{triple}
\end{equation} 
$$\overline{\widetilde{\boldsymbol{q}}+\boldsymbol{q}'}=0,\ \langle \boldsymbol{q}\rangle=\overline{\boldsymbol{q}} +\widetilde{\boldsymbol{q}},\ \langle\boldsymbol{q'}\rangle=0. $$
Substituting the decomposition into the governing equations results in a coupled system of equations for the mean flow and the coherent fluctuations and omitting the terms $\big \{ \mathcal{O}((q_i-\overline{q}_i)(q_j-\overline{q}_j))\big| \ i=1,2,3,4, \ \ j=1,3,4 \big \}$ (i.e. omitting all terms involving products of two or more fluctuating quantities except for terms involving products of up to two velocity fluctuation components), which is a reasonable approximation for low-Mach number flows considering that fluctuations of density and temperature vanish in the incompressible limit, the LNSE take the form:
\begin{subequations}
    \begin{equation}
        \grad \cdot(\overline{\rho}\overline{\boldsymbol{u}})= 0
     \end{equation}
    \begin{equation}
            \partial_t\widetilde{\rho} + L_C(\overline{\rho},\overline{\boldsymbol{u}})\widetilde{\rho}=0
    \end{equation}
    \label{couple1}
\end{subequations}
\begin{subequations}
\begin{equation}
     \boldsymbol{N}(\overline{\varrho},\overline{\boldsymbol{u}})=-\grad \cdot (\overline{\rho} \overline{\widetilde{\boldsymbol{u}} \widetilde{\boldsymbol{u}}} + \overline{\rho}\overline{\boldsymbol{u}' \boldsymbol{u}'} ) 
   \end{equation}
    \begin{equation}       \overline{\varrho}\partial_t\widetilde{\boldsymbol{u}}+\boldsymbol{L}_N(\overline{\rho},\overline{\boldsymbol{u}})\widetilde{\boldsymbol{u}}= -\grad \cdot (\overline{\rho} \widetilde{\widetilde{\boldsymbol{u}} \widetilde{\boldsymbol{u}}} + \overline{\rho}\widetilde{\boldsymbol{u}' \boldsymbol{u}'} )
\end{equation}
\end{subequations}
\begin{subequations}
    \begin{equation}
        \boldsymbol{E}(\overline{\rho},\overline{\boldsymbol{u}},\overline{T})=- \gamma(\gamma-1) \frac{M^2}{\text{Re}}\bigg[\overline{\boldsymbol{\tau}(\boldsymbol{u}'):\boldsymbol{d}(\boldsymbol{u}')} + \overline{\boldsymbol{\tau}(\widetilde{\boldsymbol{u}}):\boldsymbol{d}(\widetilde{\boldsymbol{u}})}\bigg]
    \end{equation}
    \begin{equation}
    \hspace{3.5cm}
    \hspace{-2cm} \overline{\varrho}\partial_t \widetilde{T} +L_E(\overline{\rho},\overline{\boldsymbol{u}},\overline{T})\widetilde{T}= - \gamma(\gamma-1) \frac{M^2}{\text{Re}}[\widetilde{\boldsymbol{\tau}(\widetilde{\boldsymbol{u}}):\boldsymbol{d}(\widetilde{\boldsymbol{u}})}
         + \widetilde{\boldsymbol{\tau}(\boldsymbol{u}'):\boldsymbol{d}(\boldsymbol{u}')}]
    \end{equation}
\end{subequations}
\begin{subequations}
\begin{equation}
    \overline{P} - \overline{\varrho} \overline{T} = 0
\end{equation}
 \begin{equation}
     \widetilde{P}-\widetilde{\rho}\overline{T}-\overline{\rho}\widetilde{T}=0
\end{equation}
\label{couple2}
\end{subequations}
The expressions for the linear operators $L_C$, $\boldsymbol{L}_N$ and $L_E$ can be found in appendix A.
The steady mean flow equations and the equations governing the evolution of the coherent fluctuations can be rewritten compactly using solution vector notation. Grouping the right hand sides into the non-linear forcing terms $[\overline{\boldsymbol{F}}_{\text{turb}}(\boldsymbol{q}'),\ \widetilde{\boldsymbol{F}}_{\text{turb}}(\boldsymbol{q}')]^{\intercal}$ and $[\overline{\boldsymbol{F}}_{\text{co}}(\widetilde{\boldsymbol{q}}), \ \widetilde{\boldsymbol{F}}_{\text{co}}(\widetilde{\boldsymbol{q}})]^{\intercal}$ which cover the effects of turbulent and coherent fluctuations respectively, and introducing the triple decomposition into equations \eqref{couple1} - \eqref{couple2} yields:
\begin{subequations}
\begin{equation}
    \boldsymbol{N}_{\text{tot}}(\overline{\boldsymbol{q}})=\overline{\boldsymbol{F}}_{\text{co}}(\widetilde{\boldsymbol{q}})+\overline{\boldsymbol{F}}_{\text{turb}}(\boldsymbol{q}')
\end{equation}
 \begin{equation}
    \partial_t\boldsymbol{B}(\overline{\boldsymbol{q}}) \boldsymbol{\widetilde{q}} +\boldsymbol{L}_{\text{tot}} (\overline{\boldsymbol{q}},\widetilde{\boldsymbol{q}})\widetilde{\boldsymbol{q}}=\widetilde{\boldsymbol{F}}_{\text{co}}(\widetilde{\boldsymbol{q}})+\widetilde{\boldsymbol{F}}_{\text{turb}}(\boldsymbol{q}')
\end{equation}
\label{compsys}
\end{subequations}  

Where $\boldsymbol{B}(\overline{\boldsymbol{q}})=\text{diag}(1,\overline{\rho},\overline{\rho}^2,0)$ is the mass matrix of the system and $\boldsymbol{L}_{\text{tot}}=[L_c,\boldsymbol{L}_N,L_E]^{\intercal}$.

\label{turbV}

The effects of turbulence on the mean flow ($\overline{\boldsymbol{F}}_{\text{turb}}(\boldsymbol{q}')$) are captured by a $k-\omega$ turbulence model which introduces the turbulent viscosity (TV) $\mu_t=\rho k/\omega$ where $k$ is the turbulent kinetic energy and $\omega$ the specific turbulent dissipation rate. The mean Reynolds stresses $\overline{\boldsymbol{u}'\boldsymbol{u}'}$ are linked to the mean traceless strain rate (Boussinesq approximation) through:  

\begin{equation}
    -\rho\overline{\boldsymbol{u}'\boldsymbol{u}'}=\mu_t(\boldsymbol{x})\left[ \grad\overline{\boldsymbol{u}} + (\grad\overline{\boldsymbol{u}})^{\intercal} -\frac{2}{3}(\grad\cdot\overline{\boldsymbol{u}})\boldsymbol{I}\right] -\frac{2}{3}\rho \overline{k} \boldsymbol{I}
\end{equation}

To link the coherent component of the turbulent Reynolds stresses $\widetilde{\boldsymbol{u}'\boldsymbol{u}'}$ to the traceless strain rate from the coherent fluctuations, the Boussinesq approximation can be used as in the work of \citet{yim2019self} and \cite{saturation_boujo_2018}:

\begin{equation}
    -\rho\widetilde{\boldsymbol{u}'\boldsymbol{u}'}=\mu_t(\boldsymbol{x})\left[\grad\widetilde{\boldsymbol{u}} + (\grad\widetilde{\boldsymbol{u}})^{\intercal} -\frac{2}{3}(\grad\cdot\widetilde{\boldsymbol{u}})\boldsymbol{I}\right] -\frac{2}{3}\rho \widetilde{k} \boldsymbol{I}.
\end{equation}

The governing equations \ref{compsys} can then be recast as:
\begin{subequations}
\begin{equation}
    \boldsymbol{N}_{\text{tot}}(\overline{\boldsymbol{q}})=\overline{\boldsymbol{F}}_{\text{co}}(\widetilde{\boldsymbol{q}}),
\end{equation}
 \begin{equation}
    \partial_t\boldsymbol{\widetilde{q}} +\boldsymbol{L}_{\text{tot}} (\overline{\boldsymbol{q}},\widetilde{\boldsymbol{q}})=\widetilde{\boldsymbol{F}}_{\text{co}}(\widetilde{\boldsymbol{q}}),
    \label{Ltv}
\end{equation}
\label{compsys2}
\end{subequations}

where in the final form of the operators $\boldsymbol{N}_{\text{tot}}(\overline{\boldsymbol{q}})$ and $\boldsymbol{L}_{\text{tot}} (\overline{\boldsymbol{q}},\widetilde{\boldsymbol{q}})$ in equations \eqref{compsys2} the modified eddy viscosity $\mu +\mu_t$ has been incorporated ($\mu=\rho_0\nu$ is the dynamic viscosity). The turbulent viscosity is obtained from an incompressible RANS  with $k-\omega$ closure alongside the mean flow variable fields $\overline{\boldsymbol{q}}$. These fields are then used to construct  $\boldsymbol{L}_{\text{tot}}$. 

\subsection{Boundary conditions and mean flow computation}

     The walls $\Gamma_\text{wall}$ are thermally insulated with no-slip velocity ($\boldsymbol{u}=0$) condition. On the outlet boundary $\Gamma_\text{out}$, the stress-free condition $[(p+(2/3)\mu\operatorname{tr}(\grad\cdot\boldsymbol{u})-2\mu\boldsymbol{S}]\cdot\boldsymbol{\hat{x}}=0$ is imposed on the velocity field. To reduce computational expense, a zero shear stress condition $\boldsymbol{u}\cdot\hat{\boldsymbol{y}}=0$ is set at the lower boundary $\Gamma_\text{slip}$, which does not affect the aeroacoustic dynamics of the side branch cavity on the upper side of the channel.
    
     In order to replicate the experimental conditions of \citet{Bourquard_Faure-Beaulieu_Noiray_2021} the inlet and outlet should not reflect waves when the equations for the fluctuations \eqref{Ltvlin} are solved. These anechoic conditions are achieved by combining suitable boundary conditions at $\Gamma_\text{in}$ and $\Gamma_\text{out}$ and buffer zones damping wave propagation. Non-reflecting boundary conditions has been the topic of intense research for the simulation of compressible flows, e.g. \citet{POINSOT1992104}, \citet{Colonious}, \citet{HU20084398} and \citet{Colonious2013}. Given that the duct acts as an acoustic wave guide whose cut-on frequency for transverse mode propagation is significantly above the cavity whistling frequency and that the cavity is relatively far from the inlet and outlet boundaries, the one-dimensional plane wave approximation is used, with the following Riemann invariant wave definition: $R_\pm=u' + p'/{\rho_0(\pm1+U_0/c)}$, with $+$ denoting downstream and $-$ upstream propagating waves. Based on the work of \citet{book}, the acoustic b.c. are then set to $R_-|_{\Gamma_\text{out}}=0, \ \  R_+|_{\Gamma_\text{in}}=0$. Acoustic waves are further dampened by grid stretching and buffer zones of thickness $l_s$ that feature artificial viscosity and additional dissipative terms. 
     
     The mean flow is obtained by using a compressible two-dimensional RANS solver with a $k-\omega$ turbulence model implemented in Ansys Fluent \citep{ansys_fluent_2024}. Boundary conditions at the inlet $\Gamma_\text{in}$ are uniform stream-wise velocity $U_b$, temperature $T=300$ K and turbulence intensity $5\%$. $\Gamma_\text{out}$ is a pressure outlet and $\Gamma_\text{wall}$ is no-slip, adiabatic wall. The reader is referred to Appendix \ref{AB} for more details on mesh selection and boundary treatment.

\subsection{Eigenvalue problem for a given mean flow}
The compressible LNSE (\ref{Ltv}) are now turned into an eigenvalue problem. First, the nonlinear terms $\overline{\boldsymbol{F}}_{\text{co}}(\widetilde{\boldsymbol{q}})$ and $\widetilde{\boldsymbol{F}}_{\text{co}}(\widetilde{\boldsymbol{q}})$ are neglected and a solution is sought out in the form of an eigenmode expansion. Substituting a solution of the form $\widetilde{\boldsymbol{q}}=\widetilde{\boldsymbol{q}}(\boldsymbol{x})e^{\lambda t}+ \ c.c.$ to (\ref{Ltv}), neglecting nonlinear terms in $\widetilde{\boldsymbol{q}}$ and considering a given mean flow $\overline{\boldsymbol{q}}$ one gets:
\begin{subequations}
\begin{equation}
    \boldsymbol{N}_{\text{tot}}(\overline{\boldsymbol{q}})=0
    \label{Ntvlin}
\end{equation}
 \begin{equation}
    \lambda\widetilde{\boldsymbol{q}} +\boldsymbol{L}_{\text{tot}} (\overline{\boldsymbol{q}},\widetilde{\boldsymbol{q}})=0
    \label{Ltvlin}
\end{equation}
\label{compsyslin}
\end{subequations}

The above eigenvalue problem associated with the LNSE is solved with a finite element method based on the open-source software FreeFEM \citep{FreeFEM}. The mesh of $N_E=42521$ triangular elements is constructed using FreeFem meshing tools (see appendix B for details and mesh convergence study). A hybrid interpolation approach is employed, using a combination of P2 elements for velocity and P1 elements for state variables. After discretization of the LNSE system, its eigenvalues and eigenmodes are completed by the shift-invert method.
\begin{figure}
    \centering
    \psfrag{sigma rad/s}[b][]{Growth rate $\sigma$ (rad s$^{-1})$}
    \psfrag{0}[][][0.8]{$0$}
    \psfrag{-10}[][r][0.8]{$(a)$}
    \psfrag{-20}[][r][0.8]{$(b)$}
    \psfrag{200}[][][0.8]{$200$}
    \psfrag{1}[r][r][0.8]{$-1000$}
    \psfrag{400}[][][0.8]{$400$}
    \psfrag{2}[r][r][0.8]{$-2000$}
    \psfrag{9}[][l][0.8]{$-1200$}
    \psfrag{3}[][][0.8]{$300$}
    \psfrag{6}[][l][0.8]{$-600$}
    \psfrag{8}[][l][0.8]{$-1800$}
    \psfrag{600}[][][0.8]{$600$}
    \psfrag{800}[][][0.8]{$800$}
    \psfrag{7}[][l][0.8]{$-10^2$}
    \psfrag{0}[][  ][0.8]{$0$}
    \psfrag{Frequency(Hz)}[][]{Frequency (Hz)}
    \includegraphics[width=0.8\linewidth]{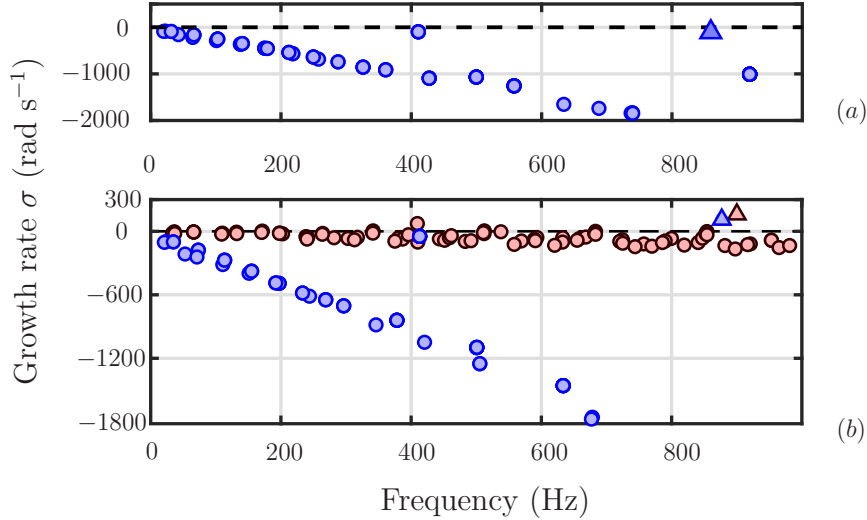}
    \caption{Eigenvalue spectrum of the compressible LNSE problem. $(a)$ $U_b = 55 \text{ m s}^{-1}$ and $H = 25$ cm
with TV. Eigenmodes are linearly stable including the 3/4 wave mode (blue triangle). $(b)$ Spectrum obtained with TV (blue) superimposed on spectrum calculated without TV (red) for the RANS mean flow obtained with $U_b = 65 \text{ m s}^{-1}$ and $H = 25$ cm.}
    \label{double}
\end{figure}

\begin{figure}
    \centering
    \psfrag{x}[][][0.7]{$x$(m)}
    \psfrag{y}[b][][0.7]{$y$ (m)}
  \psfrag{ux}[t][r][1]{$\widetilde{\boldsymbol{u}}_x$}
  \psfrag{py}[][t][1]{$\widetilde{\boldsymbol{u}}_y$}
  \psfrag{p}[][t][0.7]{$\widetilde{P}$}
  \psfrag{0.03}[r][r b][0.7]{$0.03$}
  \psfrag{-0.03}[r][r][0.7]{$-0.03$}
  \psfrag{0.05}[r][r][0.7]{$0.05$}
  \psfrag{0.08}[r][r][0.7]{$0.08$}
  \psfrag{0.058}[][t][0.6]{$(a)$}
  \psfrag{0.048}[][t][0.6]{$(b)$}
   \psfrag{0.038}[][t][0.6]{$(c)$}
    \psfrag{0.028}[][t][0.6]{$(d)$}
  \psfrag{0}[][r][0.7]{$0$}
  \psfrag{1}[][][0.7]{$5.5$}
  \psfrag{2}[][][0.7]{$-5.5$}
  \psfrag{ux}[b][tr][1]{$\widetilde{\boldsymbol{u}}_x$}
  \psfrag{0.03}[][l][0.7]{$0.03$}
  \psfrag{-0.03}[][l][0.7]{$-0.03$}
  \psfrag{0.05}[][l][0.7]{$0.05$}
  \psfrag{0.08}[][l][0.7]{$0.08$}
  \psfrag{0.3}[][l][0.7]{$0.3$}
  \psfrag{0.2}[][l][0.7]{$0.2$}
  \psfrag{0.1}[][l][0.7]{$0.1$}
  \psfrag{-0.06}[r][r][0.7]{$0.06$}
  \psfrag{P+}[l][][0.7]{$50$}
  \psfrag{P-}[l][][0.7]{$-50$}
  \psfrag{P0}[][][0.7]{$0$}
  \psfrag{Pa}[l][r][0.7]{(Pa)}
  \psfrag{ms}[t][b][0.6]{(m s$^{-1}$)}
  \includegraphics[width=\linewidth]{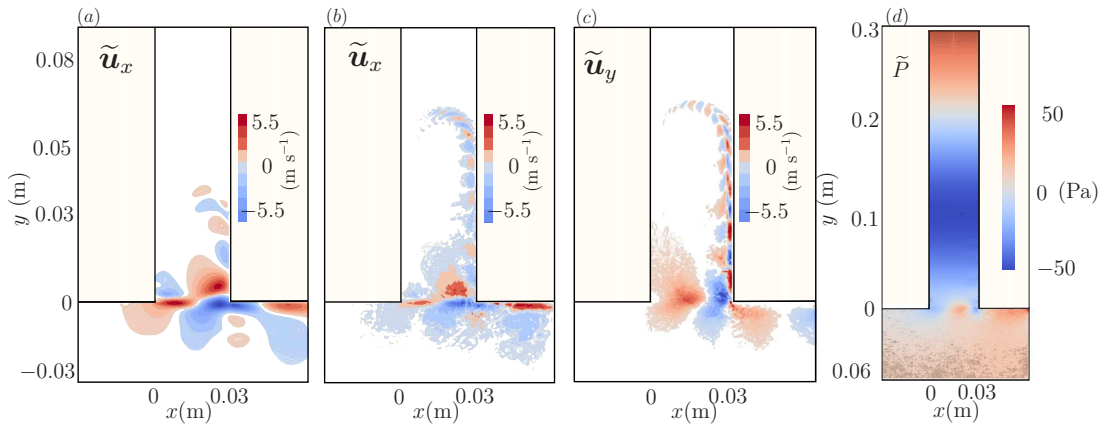}
  \caption{Dominant eigenmode of the compressible LNSE fore the mean flow obtained with $U_b=55$ m s$^{-1}$ and $H=25$ cm. Coherent component of fluctuating velocity $(a-c)$, with TV $(a)$ and without TV $(b)$-$(c)$, and pressure $(d)$ of the 3/4 mode.}
  \label{TVnoTV}
\end{figure}

Solving the LNSE system produces a spectrum which typically exhibits a branch of stable eigenvalues as shown in  figure \ref{double}\textcolor{red}{\textit{a}} for the case of the RANS mean flow obtained with $U_b=55$ m\,s$^{-1}$. Their negative growth rate (imaginary part) is generally smaller when their frequency (real part) gets larger. In this spectrum, less stable eigenvalues can be found above the branch.  These eigenvalues correspond to eigenmodes involving intense fluctuations along the shear layer developing at the cavity opening. In figure \ref{double}\textcolor{red}{\textit{a}}, the least stable one around $900$ Hz, exhibits a fluctuating pressure field which corresponds to the $3/4$ wave eigenmode of the pure acoustic problem, and velocity fluctuations in the shear layer which correspond to the first Kelvin-Helmholtz (K-H) mode of the pure hydrodynamic, i.e. incompressible, problem. The axial component of the velocity field of this mode and its pressure field are shown in figures \ref{TVnoTV}\textcolor{red}{\textit{a}} and \ref{TVnoTV}\textcolor{red}{\textit{d}}. In contrast to the results obtained with the incompressible LNSE by \citet{saturation_boujo_2018} were K-H eigenmodes were always linearly stable, i.e. with negative growth rate, the compressible LNSE considered in this work can lead to linearly unstable, i.e. with positive growth rate, aeroacoustic eigenmodes for some ranges of cavity depth $H$ and bulk flow velocity $U_b$, in agreement with experimental observations of \citet{Bourquard_Faure-Beaulieu_Noiray_2021}.

Furthermore the TV has a strong damping effect on the eigenvalues of the compressible LNSE problem. It reduces substantially the growth rate of all modes. This is illustrated in figure \ref{double} for the case for $U_b=65$ m s$^{-1}$, $H=25$ cm treated with and without TV for the same RANS mean flow. The experiments of \citet{Bourquard_Faure-Beaulieu_Noiray_2021} show that the supercritical Hopf bifurcation which separates cavity resonance condition from aeroacoustic limit cycles occurs close to this bulk flow velocity $U_b$. Comparing figures \ref{TVnoTV}\textcolor{red}{\textit{b}} and \ref{TVnoTV}\textcolor{red}{\textit{c}} one can see that TV also modifies the mode shape by reducing the penetration depth of the fluctuation inside the cavity. All the results presented in the remainder of this paper are obtained with TV. Moreover, for all combinations of $U_b$ and $H$ considered in this work, one mode dominates the spectrum of eigenvalues of the compressible LNSE associated to the RANS mean flow. It corresponds to the 3/4 mode of the cavity with an acoustic pressure wavelength equal to approximately 4$H$/3.

\subsection{Self-consistent saturation model}
\label{selfsat}
For cases where the compressible NS equations linearized around the RANS mean flow lead to a linearly unstable $3/4$ wave eigenmode, infinitesimal perturbations exponentially grow and an acoustic limit cycle is ultimately reached. The method to predict its amplitude is now explained. It is based on the self-consistent analysis of hydrodynamic limit cycles proposed by \citet{PhysRevLett.113.084501}. They validated their method with the incompressible laminar self-oscillating  wake behind a cylinder and developed it by building upon the work of \citet{Stuart_1958}. Assuming that the mode shape does not change with the amplitude, \cite{Stuart_1958} solved an amplitude dependent LNSE eigenvalue problem by subtracting a scaled mean component of the coherent Reynolds stress divergence such that a marginally stable base flow solution is found. The scaling factor then provides an estimate of the stable limit cycle amplitude. However this approach often produces an incorrect estimate of the amplitude because the shape of the mode exerting the steady forcing on the mean flow usually depends on the oscillation amplitude. An elegant way to overcome this issue was proposed by \citet{PhysRevLett.113.084501}. The saturated state is iteratively obtained by gradually increasing the value of the forcing amplitude and resolving the updated equations at each step. It artificially creates a feedback loop, where steady forcing from the unstable mode (see an example in figure \ref{ModeRS}) modifies the mean flow, which in turn lead to a modified mode shape and so on until the mode amplitude increment leads to a marginally stable mode. This development mimics the physical feedback mechanism that is responsible for saturation processes by gradually varying the shape of the base flow (and the unstable mode), leading to accurate and robust convergence behavior.
As mentioned in the introduction, the self-consistent saturation analysis performed in this study account for both the effects of turbulence and compressibility. Turbulence effects are accounted for through the terms originating from the triple decomposition of the flow field and an eddy-viscosity closure model as explained in the previous sections, while the effects of compressibility are resolved by the compressible equations for density, temperature and pressure in addition to the momentum equation. The resulting system of equations describing the self-consistent saturation model, including all nonlinear terms up to first order, can be found in appendix \ref{AC} or in the work of \citet{AcousticCyl}. For the present low-Mach flow cases, one of the terms dominates the others by more than an order of magnitude and is the only one retained in the present model. Thus, the resulting system in solution-vector notation reads: 
\begin{figure}
    \centering
          \psfrag{y}[b][][0.7]{$y$(m)}
      \psfrag{X}[][l][0.7]{$x$(m)}
      \psfrag{0}[][][0.7]{$0$}
      \psfrag{1}[][][0.7]{$1$}
      \psfrag{2}[][][0.7]{$2$}
      \psfrag{3}[][][0.7]{$3$}
      \psfrag{4}[][][0.7]{$4$}
      \psfrag{5}[][][0.7]{$5$}
      \psfrag{6}[][][0.7]{$6$}
      \psfrag{30}[][][0.7]{$30$}
      \psfrag{25}[][][0.7]{$25$}
      \psfrag{20}[][][0.7]{$20$}
      \psfrag{15}[][][0.7]{$15$}
      \psfrag{10}[][][0.7]{$10$}
      \psfrag{5}[][][0.7]{$5$}
      \psfrag{-5}[][l][0.7]{$-5$}
      \psfrag{a}[][l][0.7]{$(a)$}
      \psfrag{b}[][l][0.7]{$(b)$}
      \psfrag{U}[l][l][0.55]{$U_b=65$ m s$^{-1}$}
      \psfrag{H}[l][l][0.55]{$H=25$ cm}
      \psfrag{ms}[][t][0.6]{(kg m$^{-2}$ s$^{-2}$)}
      \psfrag{iteration}[][b][0.7]{iteration}
      \psfrag{Growth rate}[][][0.7]{Growth rate $\sigma$ (rad s$^{-1}$)}
      \psfrag{RSx}[l][][0.8]{$\Re\{(\overline{\rho}\boldsymbol{\widetilde{u}}^*\cdot\grad\boldsymbol{\widetilde{u}})\cdot\hat{x}\}$}
      \psfrag{RSy}[l][][0.8]{$\Re\{(\overline{\rho}\boldsymbol{\widetilde{u}}^*\cdot\grad\boldsymbol{\widetilde{u}})\cdot\hat{y}\}$}
      \psfrag{0.03}[][l][0.7]{$0.03$}
      \psfrag{40}[l][][0.6]{$3\times10^6$}
      \psfrag{-40}[l][][0.6]{$-3\times10^6$}
      \psfrag{-0.03}[][l][0.7]{$-0.03$}
     \psfrag{1}[l ][r][0.7]{$1$}
     \psfrag{-1}[][r][0.7]{$-1$}
    \includegraphics[width=\linewidth]{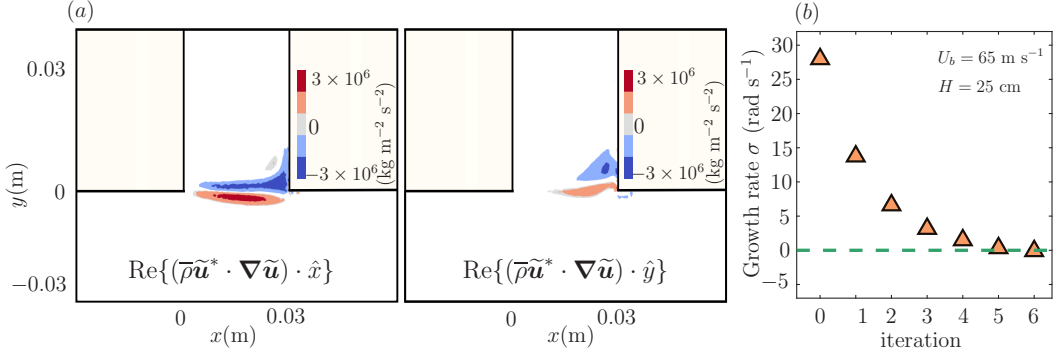}
    \caption{$(a)$ Horizontal and vertical components of the steady forcing from the Reynolds stress of the coherent fluctuations for $U_b=65$ m s$^{-1}$ and $H=25$ cm at the first iteration. $(b)$ Evolution of the linear growth rate of the unstable aeroacoustic mode during the 6 steps of the self consistent iterative process.}
    \label{ModeRS}
\end{figure}
\begin{subequations}
\begin{equation}
    \boldsymbol{N}_{\text{tot}}(\overline{\boldsymbol{q}}^{(j+1)})=-2A^2\Re\{\overline{\rho}\widetilde{\boldsymbol{u}}_{\text{un}}^{(j)*}\cdot\grad\widetilde{\boldsymbol{u}}_{\text{un}}^{(j)}\}\cdot[0,\boldsymbol{1},0,0]^{\intercal},
\label{satsys1}
\end{equation}
 \begin{equation}
    \lambda\widetilde{\boldsymbol{q}}^{(j)} +\boldsymbol{L}_{\text{tot}} (\overline{\boldsymbol{q}}^{(j)} ,\widetilde{\boldsymbol{q}}^{(j)} )=0
\label{satsys2}
\end{equation}
\label{satsys}
\end{subequations} 
where $j\in\mathbb{N}^*$ is the iteration step and $\widetilde{\boldsymbol{u}}_{\text{un}}^{(j)}$ is the velocity component of the unstable eigenmode $\widetilde{\boldsymbol{q}}^{(j)}_{\text{un}}$, and $\widetilde{\boldsymbol{u}}_{\text{un}}^{(j)*}$ its complex conjugate. The first mean flow $\overline{\boldsymbol{q}}^{(1)}$ used to obtain $\widetilde{\boldsymbol{q}}^{(1)}_{\text{un}}$ with equation \eqref{satsys2} is the RANS solution. The forcing	term on the RHS of equation \eqref{satsys1} is presented in figure \ref{ModeRS}\textcolor{red}{\textit{a}}  for $U_b=65$ m\,s$^{-1}$ and $H=25$ cm. In figure \ref{ModeRS}\textcolor{red}{\textit{b}}, one can see the evolution of the linear growth rate of the unstable mode during the iteration process leading to marginal stability after six steps.  The constant $A$ in equation \eqref{satsys1} can be interpreted as the mode amplitude, and as proposed by \citet{PhysRevLett.113.084501} in their algorithm, its value is determined by requiring the value of $A_f^2=A^2 \Vert{2\Re\{\overline{\rho}\widetilde{\boldsymbol{u}}_{\text{un}}^{(j)*}\cdot\grad\widetilde{\boldsymbol{u}}_{\text{un}}^{(j)}}\Vert/\Vert{\overline{\rho}\widetilde{\boldsymbol{u}}_{\text{un}}^{(j)}}\Vert^2$ to be fixed for ensuring convergence of the iteration process. Introducing $A_f$ is not strictly necessary for the iterative process. Its use explicitly shows the normalization of the eigenvectors using the metric of the space over which they are defined. In the present discretized LNSE operator space the dot product of two vectors is defined using the mass matrix of the problem $\boldsymbol{B}(\overline{\boldsymbol{q}})$ with $\Vert{\tilde{\boldsymbol{q}}}\Vert^2=\tilde{\boldsymbol{q}}^{\dagger}\boldsymbol{B}(\overline{\boldsymbol{q}})\tilde{\boldsymbol{q}}\simeq \iint_\Omega\langle\overline{\rho}\tilde{u}_i^*\tilde{u}_i\rangle \text{d}\Omega=1$. 
\begin{figure}
    \centering
    \psfrag{S}[][l][0.7]{Growth rate $\sigma$ (rad s$^{-1}$)}
    \psfrag{iteration}[][l][0.7]{Iterations}
    \psfrag{s}[t][l][0.7]{Final growth rate $\sigma$ (rad s$^{-1}$)}
    \psfrag{Af}[][r][0.99]{$A_f$}
    \psfrag{15}[][][0.7]{$15$}
    \psfrag{5}[][][0.7]{$5$}
    \psfrag{50}[][][0.7]{$50$}
    \psfrag{100}[][][0.7]{$100$}
    \psfrag{0}[][b][0.7]{$0$}
    \psfrag{5}[][][0.7]{$5$}
    \psfrag{10}[][l][0.7]{$10$}
    \psfrag{0.26}[][][0.7]{$0.25$}
    \psfrag{0.29}[][][0.7]{$0.29$}
    \psfrag{0.3}[][][0.9]{$0.3$}
    \psfrag{a1}[][][0.9]{$(a)$}
    \psfrag{a2}[][][0.9]{$(b)$}
    \psfrag{a3}[][][0.9]{$(c)$}
    \psfrag{b}[][][0.9]{$(d)$}
    \includegraphics[width=0.85\linewidth]{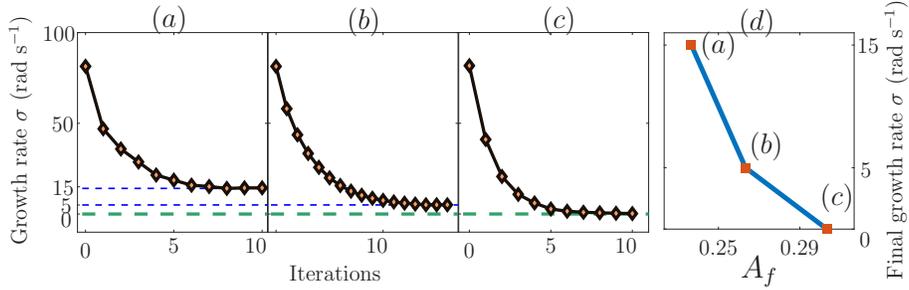}
    \caption{$(a-c)$ Convergence of the growth rate during the self-consistent model iterations for different values of $A_f$ and for $U_b=68$ m s$^{-1}$ and $H=25$ cm. $(d)$ Final growth rate for monotonically increasing $A_f$. The marginal stability, i.e. $\sigma=0$ rad s$^{-1}$, is obtained for $A_f=0.3$ in this case. }
    \label{roleAf}
\end{figure}
An adequate tuning of $A_f$ is needed for ensuring that a marginally stable (vanishing growth rate) is reached and for minimizing the number of steps (see an example in figure \ref{roleAf}). Following \citet{PhysRevLett.113.084501}, under-relaxation is also used to ensure robust convergence by updating the mean flow with $\overline{\boldsymbol{u}}^{(j+1)}=\gamma \overline{\boldsymbol{u}}^{(j-1)} +(1 - \gamma) \overline{\boldsymbol{u}}^{(j)} , \ \text{for }j\geq2 $ where $\gamma\in(0,1)$ is the under-relaxation factor. 
\section{Results}
\begin{figure}[t!]
    \centering
      \psfrag{y}[br][tl][0.7]{$y$ (m)}
      \psfrag{x}[][][0.7]{$x$ (m)}
      \psfrag{ms}[][t][0.5]{(m s$^{-1}$)}
       \psfrag{10}[l][r][0.6]{$0.2$}
       \psfrag{20}[l][r][0.6]{$0.5$}
       \psfrag{30}[l][r][0.6]{$0.7$}
       \psfrag{40}[l][r][0.6]{$0.9$}
       \psfrag{50}[l][r][0.6]{$1.1$}
       \psfrag{60}[l][r][0.6]{$1.6$}
       \psfrag{-10}[l][][0.6]{$-0.2$}
       \psfrag{-20}[l][][0.6]{$-0.5$}
       \psfrag{-30}[l][][0.6]{$-0.7$}
       \psfrag{-40}[l][][0.6]{$-0.9$}
       \psfrag{-50}[l][][0.6]{$-1.1$}
       \psfrag{-60}[l][][0.6]{$-1.6$}
       \psfrag{00}[l][r][0.6]{$0$}
      \psfrag{0}[][][0.6]{$0$}
      \psfrag{RSx}[l][][0.8]{$\widetilde{\boldsymbol{u}}_x$}
      \psfrag{3}[][l][0.6]{$0.03$}
      \psfrag{-3}[][l][0.6]{$-0.03$}
     \psfrag{1}[][][0.7]{$1$}
     \psfrag{-1}[][l][0.7]{$-1$}
      \psfrag{i0}[l][][0.5]{iteration $0$}
      \psfrag{i1}[l][][0.5]{iteration $1$}
      \psfrag{i3}[l][][0.5]{iteration $2$}
      \psfrag{i4}[l][][0.5]{iteration $3$}
      \psfrag{i5}[l][][0.5]{iteration $4$}
      \psfrag{i6}[l][][0.5]{iteration $5$}
    \includegraphics[width=0.8\linewidth]{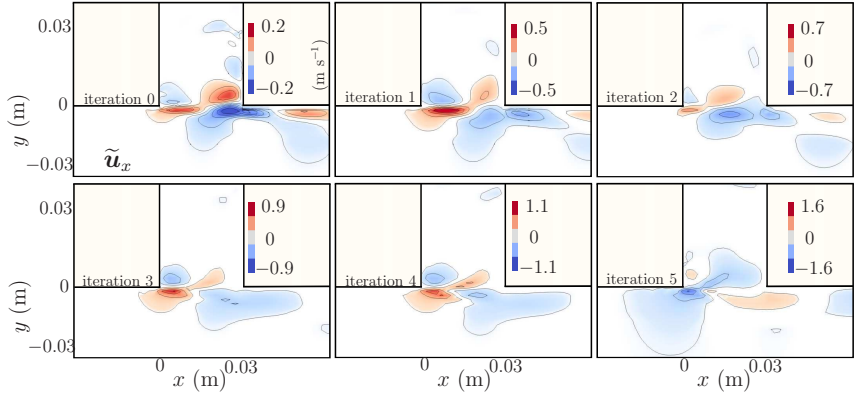}
    \caption{Evolution of stream wise velocity of the unstable mode over the course of the iterative process for $U_b=65$ m s$^{-1}$ and $H=25$ cm.}
    \label{mode_evo}
\end{figure}
\begin{figure}[t!]
    \centering
      \psfrag{y}[br][tl][0.7]{$y$ (m)}
      \psfrag{x}[][][0.7]{$x$ (m)}
      \psfrag{ms}[][t][0.5]{(kg m$^{-2}$ s$^{-2}$)}
       \psfrag{40}[l][][0.5]{$3\times10^6$}
      \psfrag{-40}[l][][0.5]{$-3\times10^6$}
      \psfrag{0}[][][0.6]{$0$}
      \psfrag{RSx}[l][][0.8]{$\Re\{(\overline{\rho}\boldsymbol{\widetilde{u}}^*\cdot\grad\boldsymbol{\widetilde{u}})\cdot\hat{x}\}$}
      \psfrag{3}[][l][0.6]{$0.03$}
      \psfrag{-3}[][l][0.6]{$-0.03$}
      \psfrag{i0}[l][][0.5]{iteration $0$}
      \psfrag{i1}[l][][0.5]{iteration $1$}
      \psfrag{i3}[l][][0.5]{iteration $2$}
      \psfrag{i4}[l][][0.5]{iteration $3$}
      \psfrag{i5}[l][][0.5]{iteration $4$}
      \psfrag{i6}[l][][0.5]{iteration $5$}
    \includegraphics[width=0.8\linewidth]{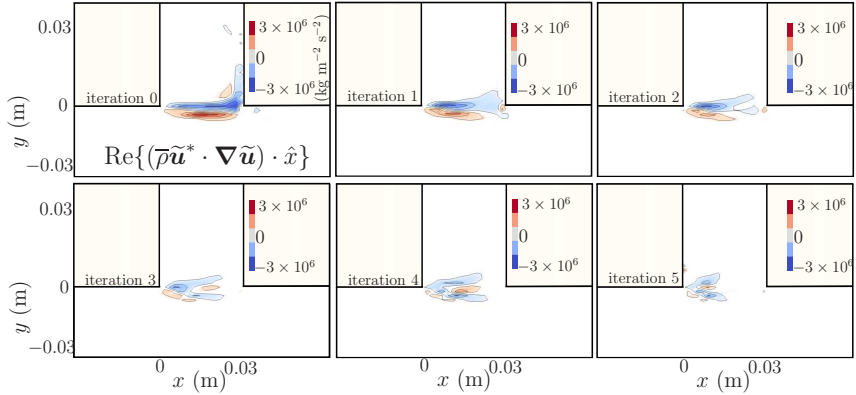}
    \caption{Evolution of stream wise Reynolds stress divergence of the unstable mode (forcing term of the mean flow equations) over the course of the iterative process for $U_b=65$ m s$^{-1}$ and $H=25$ cm.}
    \label{mode_evo_RSx}
\end{figure}

\begin{figure}
    \centering
    \psfrag{U=72}[l][l][0.6]{\color[RGB]{0, 81, 13}{$U_b=72$ m s$^{-1}$}}
    \psfrag{H=25}[l][l][0.6]{\color[RGB]{0, 53, 81}{$H=25$ cm}}
    \psfrag{Amp}[b][][0.9]{$A$ (m s$^{-1}$)}
    \psfrag{0}[][][0.7]{$0$}
    \psfrag{1}[][][0.7]{$1$}
    \psfrag{2}[][][0.7]{$2$}
    \psfrag{5}[][][0.7]{$5$}
    \psfrag{10}[][][0.7]{$10$}
    \psfrag{15}[][l][0.7]{$15$}    
    \psfrag{Fdif}[][][0.9]{$\abs{\Delta f}$ \ (Hz)}
    \psfrag{ImZ}[t][][0.9]{$\Im(\mathcal{Z})$}
    \psfrag{0.5}[][r][0.7]{$0.5$}  
    \psfrag{800}[][][0.7]{$800$}
    \psfrag{1000}[][][0.7]{$1000$}
    \psfrag{1200}[][][0.7]{$1200$}
    \psfrag{20}[][][0.7]{$20$}
    \psfrag{50}[][][0.7]{$50$}
    \psfrag{100}[][][0.7]{$100$}
    \psfrag{200}[][][0.7]{$200$}
    \psfrag{400}[][][0.7]{$400$}
    \psfrag{Pa}[][][0.7]{Pa}
    \psfrag{f}[][][0.9]{$f$ (Hz)}    
    \includegraphics[width=0.75\linewidth]{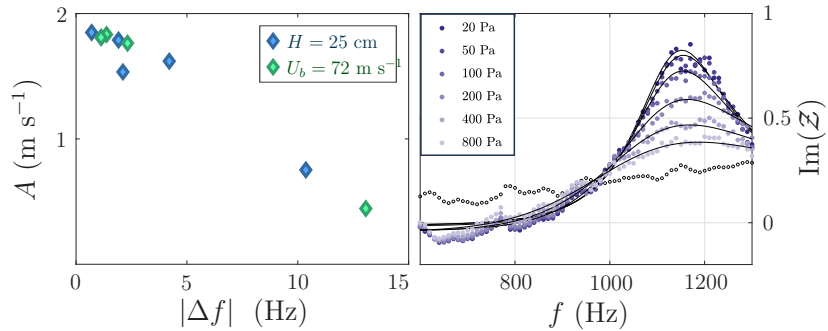}
    \caption{On the left, the correlation of converged self-oscillating amplitude and resulting frequency correction obtained from the self-consistent saturation model. On the right, the reactance of the cavity opening for different acoustic
forcing amplitudes and for $U_b= 74$ m\,s$^{-1}$ measured and presented by \cite{Bourquard_Faure-Beaulieu_Noiray_2021}. The white circles correspond to the specific impedance without flow.}
    \label{FreqCorr}
\end{figure}

The self-consistent model is now employed for the case of $U_b=65$ ms$^{-1}$ and $H=25$ m.  The initial step leads to a forcing from the steady component of the coherent Reynolds stress presented in figure \ref{ModeRS}\textcolor{red}{\textit{a}}, and to a linear growth rate of approximately $30$ rad s$^{-1}$ (see figure \ref{ModeRS}\textcolor{red}{\textit{b}}). Figures \ref{mode_evo} and \ref{mode_evo_RSx} contain snapshots of streamwise fluctuation velocity and Reynolds stress terms respectively for the first 6 iterations. Figure \ref{mode_evo_RSx}  clearly demonstrates the decay of the steady Reynolds stress forcing magnitude, when approaching saturation. This decrease in amplitude signifies a reduction in energy transfer between coherent fluctuations and mean flow, hinting at an energetic equilibrium state that will be reached upon complete saturation. Furthermore, one can see in figures \ref{mode_evo} and \ref{mode_evo_RSx} that the region exhibiting the largest coherent fluctuating velocity and the associated steady Reynolds stress forcing progressively migrates toward the upstream corner from an initial distribution spread across the cavity opening. The prediction of this effect from the self-consistent analysis based on the compressible LNSE is in excellent agreement with the compressible Large Eddy Simulations presented by \citet{saturation_boujo_2018}. Indeed, in their work, coherent velocity fluctuations are imposed at the upper end of the side branch and the forcing amplitude is progressively increased, resulting in a very similar modification of the coherent response and of their steady Reynolds stress component. Interestingly, the region toward which high amplitude coherent fluctuations migrate during the self-consistent saturation process is characterized by a high sensitivity to forcing. Indeed, although the steady Reynolds stress forcing amplitude decreases, the effect it has on the mean flow increases. This result is in agreement with the adjoint-based sensitivity analysis of the incompressible flow from  \citet{saturation_boujo_2018}. Close to the upstream corner of the cavity, small perturbations are strongly amplified as they are convected downstream along the convectively unstable shear layer. This mechanism leads to a natural reorganization of the mean flow and its unstable mode during the saturation process leading to stable self-oscillations of the aeroacoustic system. 
\begin{figure}
    \centering
    \psfrag{A [m/s]}[][r][0.9]{$A$ (m s$^{-1}$)}
    \psfrag{A}[l][][0.9]{$A$ (m s$^{-1}$)}
    \psfrag{G}[l][][0.9]{$\sigma$ (rad s$^{-1}$)}
    \psfrag{H=24cm}[][][0.8]{\color[RGB]{0, 53, 81}{\ $H=24$ cm}}
    \psfrag{H=25cm}[][][0.8]{\color[RGB]{0, 53, 81}{\ $H=25$ cm}}
    \psfrag{Uz=75m/s}[][l][0.8]{\color[RGB]{0, 81, 13}{$  U_b=75$  m s$^{-1}$}} 
    \psfrag{Uz=72m/s}[][l][0.8]{\color[RGB]{0, 81, 13}{$   U_b=72$  m s$^{-1}$}}     
    \psfrag{Uz}[][b][0.9]{$U_b$}
    \psfrag{H}[][b][0.9]{$H$}
    \psfrag{0}[][][0.7]{$0$}
    \psfrag{1}[][][0.7]{$1$}
    \psfrag{10}[][l][0.7]{$10$}
    \psfrag{5}[][][0.7]{$5$}
    \psfrag{91}[][l][0.7]{$91$}
    \psfrag{20}[][l][0.7]{$20$}
    \psfrag{2}[][][0.7]{$2$}
    \psfrag{3}[][][0.7]{$3$}
    \psfrag{60}[][][0.7]{$60$}
    \psfrag{75}[][][0.7]{$75$}
    \psfrag{90}[][l][0.7]{$80$}
    \psfrag{200}[][][0.7]{$200$}
    \psfrag{270}[][][0.7]{$270$}
    \psfrag{iteration}[][l][0.8]{iterations}
    \psfrag{a}[][l][0.8]{$(a)$}
    \psfrag{b}[][l][0.8]{$(b$)}
    \includegraphics[width=0.75\linewidth]{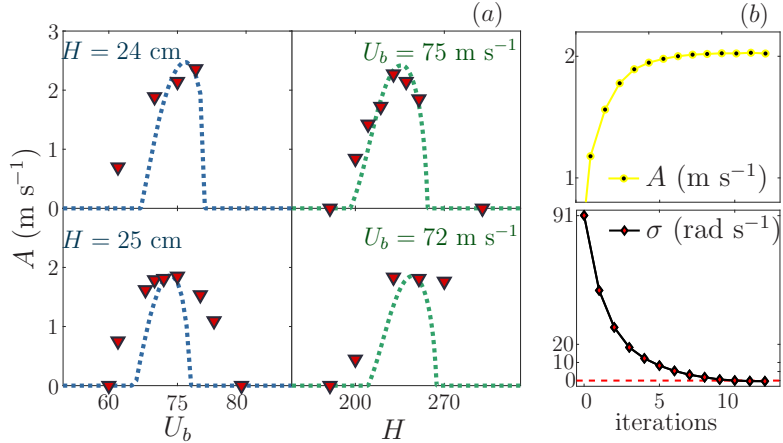}
    \caption{$(a)$ Coherent fluctuation amplitudes computed with the self consistent analysis of the compressible LNSE (red triangles) plotted over the amplitude curves of the experimentally calibrated low-order  model of  \citet{Bourquard_Faure-Beaulieu_Noiray_2021}. $(b)$ Evolution of the coherent fluctuation unstable mode amplitude (up) and growth rate (down) over the iterations of the self-consistent analysis for $U_b=80$ m s$^{-1}$ and $H=21$ cm.}
    \label{compar}
\end{figure}
As shown in figure \ref{compar}, predictions for the saturation amplitude of the limit cycle oscillations over a range of the flow velocity $U_b$ and the height of the cavity $H$ are in good agreement with the experimental data of \citet{Bourquard_Faure-Beaulieu_Noiray_2021} and with their low-order model after its parameters have been calibrated. In the high-amplitude regimes, the self-consistent saturation model converges robustly and captures the non-linear amplitude limiting mechanism attributed to a balance of acoustic energy sources and losses from the self-oscillating flow of the aperture and acoustic energy losses at the channel boundaries. However, for combinations of $U_b$ and $H$ in the vicinity of the supercritical Hopf bifurcation (for instance for $U_b=60$ m\,s$^{-1}$ and $H=25$ cm), the self-consistent model overestimates the amplitude. This discrepancy arises from the minimum number of iterations required by the self-consistent model to converge, artificially increasing the saturated state amplitude \citep{PhysRevLett.113.084501}. 
\begin{figure}
    \centering
    \psfrag{f}[][][0.6]{$f$ (Hz)}
    \psfrag{1}[][][0.6]{$H=20$ cm,  A=0.44}
    \psfrag{2}[][][0.6]{$H=23$ cm,  A=1.83}
    \psfrag{3}[][][0.6]{$H=27$ cm,  A=1.76}
    \psfrag{4}[][][0.6]{$U_b=62$ m s$^{-1}$, A=0.8}
    \psfrag{A}[][][0.6]{$U_b=75$ m s$^{-1}$, A=1.8}
    \psfrag{6}[][][0.6]{$U_b=83$ m s$^{-1}$, A=1.1}
    \psfrag{Iter}[][][0.7]{iteration}
    \psfrag{Uz=72}[][][0.7]{\color[RGB]{0, 81, 13}{$U_b=72$ m s$^{-1}$}}
    \psfrag{H=25}[][][0.7]{\color[RGB]{0, 53, 81}{$H=25$ cm}}
    \psfrag{1135}[][][0.6]{$1135$}
    \psfrag{1122}[][][0.6]{$1122$}
    \psfrag{0}[][][0.6]{$0$}
    \psfrag{5}[][][0.6]{$5$}
    \psfrag{10}[][][0.6]{$10$}
    \psfrag{15}[][][0.6]{$15$}
    \psfrag{20}[][][0.6]{$20$}
    \psfrag{25}[][][0.6]{$25$}
    \psfrag{30}[][][0.6]{$30$}
    \psfrag{981}[][][0.7]{$981$}
    \psfrag{983}[][][0.7]{$983$}
    \psfrag{893}[][][0.7]{$893$}
    \psfrag{891}[][][0.7]{$891$}
    \psfrag{939}[][][0.7]{$939$}
    \psfrag{929}[][][0.7]{$929$}
    \psfrag{954}[][][0.7]{$954$}
    \psfrag{953}[][][0.7]{$953$}    
    \psfrag{965}[][][0.7]{$965$}
    \psfrag{963}[][][0.7]{$963$}    
    \includegraphics[width=0.8\linewidth]{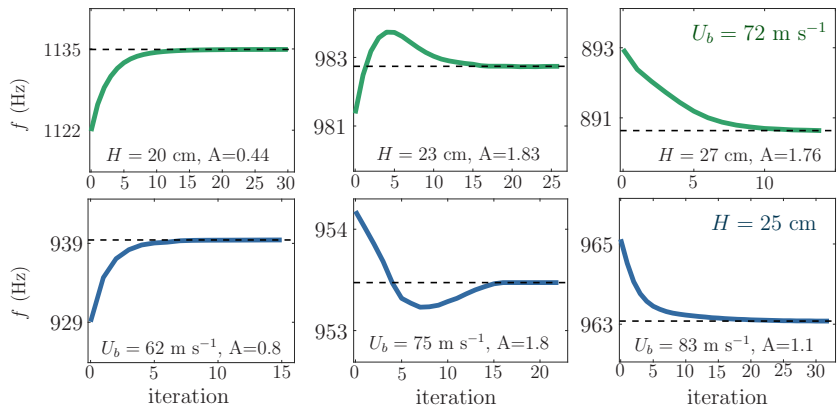}
    \caption{Frequency convergence over successive iterations of the self consistent analysis. Upper row: bulk velocity is kept constant at $U_b=72 \text{ m s}^{-1} $ and the cavity depth is varied. Lower row: The cavity depth is kept constant at $H=25$ cm and the bulk velocity is varied.}
    \label{FreqEvo}
\end{figure}
Moreover, the overestimation of the oscillation amplitude close to the Hopf points can also be attributed to overestimated linear growth rate of the mode, themselves due to underestimated aeroacoustic damping in the present 2D analysis which does not account for flow friction along the lateral boundaries of the experiment from \citet{Bourquard_Faure-Beaulieu_Noiray_2021}. The frequency corrections predicted by the model over the linear stability theory predictions exhibit a systematic and well interpretable trend. As saturation amplitude decreases, the magnitude of the frequency correction increases. The sign of the correction is negative for frequencies above the system's reactance inflection point and positive below it, in good agreement with the acoustic reactance  measured in \citet{Bourquard_Faure-Beaulieu_Noiray_2021} (see figures \ref{FreqEvo} and \ref{FreqCorr} respectively).

\begin{figure}
    \centering
    \psfrag{y}[b][][1]{$y$ (m)}
    \psfrag{x}[t][]{$x$ (m)}
    \psfrag{-0.025}[][][0.8]{$-0.025$}
    \psfrag{0.025}[][][0.8]{$0.025$}
    \psfrag{0}[][][0.8]{$0$}
    \psfrag{0.025}[][][0.8]{$0.025$}
    \psfrag{0.05}[][][0.8]{$0.05$}
    \psfrag{0.06}[][][0.8]{$0.06$}
    \psfrag{0.03}[][][0.8]{$0.03$}
    \psfrag{70}[][r][0.8]{$70$}
          \psfrag{ms}[][t][0.7]{(m s$^{-1}$)}
    \psfrag{U}[t][b]{$\overline{\boldsymbol{u}}$}
    \includegraphics[width=0.5\linewidth]{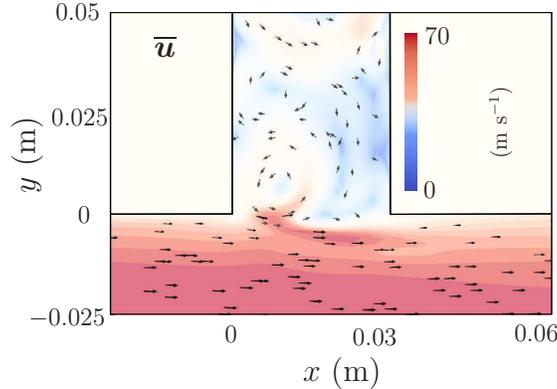}
    \caption{Converged, stable mean velocity magnitude $\overline{\boldsymbol{u}}$ for a  cavity of height $H=28$ cm and inlet bulk velocity $U_b=65$ m s$^{-1}$.}
    \label{ConvMean}
\end{figure}

Finally, it is interesting to discuss the saturated mean flow produced by self-consistent analysis. The example of the saturated mean flow for $U_b=65$ m\,s$^{-1}$ and $H=28$ cm is given in figure \ref{ConvMean}. It exhibits multiple cells that do not span accross the width of the aperture. This final mean velocity distribution leads to a marginally stable $3/4$ wave aeroacoustic mode, in agreement with the experiments.
\section{Conclusions}
The aeroacoustic limit cycles of deep cavities subjected to low-Mach turbulent flows have been predicted with a self-consistent model based on the compressible LNSE. It provides intermediate snapshots of the saturation process and make accurate predictions without the need for calibration data from simulations or experiments. Turbulent viscosity was incorporated to the self-consistent model using the triple decomposition of flow field. The predicted amplitude and frequency of the self-oscillating mode are in good agreement with the ones obtained from an experimentally calibrated low-order model, validating the capability of this framework to capture nonlinear effects beyond linear stability predictions.
Moreover, the present work shows that the self-consistent analysis reproduces key features of the self-oscillating flow, such as the amplitude dependence of the frequency shift, with corrections being more significant as the amplitude decreases, in agreement with previously published experimental results. The model tends to overpredict amplitudes in the vicinity of the supercritical Hopf bifurcation, presumably because iterative convergence criteria are less robust at small amplitude, and because the three-dimensional effects of aeroacoustic dissipation in the lateral boundary layer of the physical system are not accounted for in the present two-dimensional analysis.
Furthermore, the reorganization of the coherent fluctuations of the flow field enhance the Reynolds stress  forcing in regions of high receptivity of the flow. It demonstrates that the saturation mechanism naturally aligns with the physical sensitivity of the initial mean flow obtained by solving the RANS equations. These results consolidate the  literature dealing with self-consistent analysis of oscillating flows and demonstrate the method's capabilities in identifying dominant feedback pathways of aeroacoustic instabilities in low-Mach turbulent flows. It also shows its potential for designing control strategies and geometries for aeroacoustic systems, given that this self-consistent method can be operated at a fraction of the computational cost of codes traditionally chosen for this task.

\begin{appendix}
\section{LNSE operators $L_C$, $\boldsymbol{L}_N$, $L_E$}
\begin{subequations}
    \begin{equation}
    L_C(\overline{\rho},\overline{\boldsymbol{u}})\widetilde{\rho}=\widetilde{\rho}\grad\cdot \overline{\boldsymbol{u}} +\overline{\boldsymbol{u}}\cdot\grad\widetilde{\rho} +\widetilde{\boldsymbol{u}}\cdot\grad\overline{\rho} +\overline{\rho}\grad\cdot\widetilde{\boldsymbol{u}}
    \end{equation}   
    \begin{equation}
    \boldsymbol{L}_N(\overline{\rho},\overline{\boldsymbol{u}})\widetilde{\boldsymbol{u}}=\widetilde{\rho}(\overline{\boldsymbol{u}}\cdot\grad)\overline{\boldsymbol{u}} +\overline{\rho}(\overline{\boldsymbol{u}}\cdot\grad)\widetilde{\boldsymbol{u}} + \overline{\rho}(\widetilde{\boldsymbol{u}}\cdot\grad)\overline{\boldsymbol{u}} + \frac{1}{\gamma M^2}\grad \widetilde{P} -\frac{1}{\text{Re}}\grad\cdot\boldsymbol{\tau}(\widetilde{\boldsymbol{u}})
    \end{equation}
    $$
    \hspace{-1.7cm}L_E(\overline{\rho},\overline{\boldsymbol{u}},\overline{T}) \widetilde{T}= \widetilde{\rho}(\overline{\boldsymbol{u}}\cdot\grad)\overline{T} +\overline{\rho}(\overline{\boldsymbol{u}}\cdot\grad)\widetilde{T} + \overline{\rho}(\widetilde{\boldsymbol{u}}\cdot\grad)\overline{T}- \frac{\gamma}{\text{Pr}\text{Re}}\nabla^2\widetilde{T}
    $$
    \begin{equation}
    + (\gamma -1)\Big(\big[\widetilde{\rho}\overline{T}(\grad\cdot\overline{\boldsymbol{u}}) + \overline{\rho}\widetilde{T}(\grad\cdot\overline{\boldsymbol{u}}) +\overline{\rho}\overline{T}(\grad\cdot\widetilde{\boldsymbol{u}})\big] 
        - \gamma \frac{M^2}{\text{Re}}\big[\boldsymbol{\tau}(\widetilde{\boldsymbol{u}}):\boldsymbol{d}(\overline{\boldsymbol{u}}) + \boldsymbol{\tau}(\overline{\boldsymbol{u}}):\boldsymbol{d}(\widetilde{\boldsymbol{u}})\big]\Big) 
    \end{equation}
\end{subequations}

\section{Domain meshes and non-reflecting boundaries}
\label{AB}
   
    Table \ref{tablemesh} provides the different parameters of the meshes used to perform the convergence study of the LNSE eigenmode problem with the RANS mean flow. Mesh M7 has been adopted for all results presented throughout the paper for cases with cavity depth $H=25$ cm and versions of it with extruded (or truncated) cavities, identical otherwise to the original, are used for cases of different depths.  

    A buffer zone approach is adopted in the present work, wherein fluctuating quantities are artificially damped before reaching the domain boundary. This method effectively reduces unwanted reflections by imposing a dissipative effect on the coherent fluctuations. The term $-\beta(x,y)\cdot\widetilde{\boldsymbol{q}}(x)$ is added to the r.h.s. of the linearized equations with the damping function $\beta(x,y)$ defined as:
    \begin{subequations}
        \begin{equation}
            \beta(x,y)=0,\textrm{ \ \ \ \ \    for } x_{in}\leq x\leq x_{out}
        \end{equation}
        \begin{equation}
            \beta(x,y)=\abs{1-\frac{1}{M}}f(x_{in},x)\ \textrm{ \ \ \ for } x\leq x_{in}
        \end{equation}
        \begin{equation}
            \beta(x,y)=\abs{1+\frac{1}{M}}f(x,x_{out})\ \textrm{ for } x\geq x_{in}
        \end{equation}
    \end{subequations}
    where $f(x,y)=2\alpha(x-y)/l_s^2$, $x_{in}=-(L_1-l_s)$ and $x_{out}=(L_2-l_s)$.The constants $\alpha$ and $l_s$ determine the damping intensity and length of the buffer zone respectively. To further enhance numerical stability, following the methodology of \citet{AcousticCyl} and \citet{Yamouni_Sipp_Jacquin_2013}, a grid stretching technique is employed to enhance dissipation near the boundaries and an artificial viscosity component scaling with the mesh element size is introduced. This involves progressively coarsening the mesh near the boundaries, further mitigating wave reflections. The artificial viscosity  is defined as: $\nu_{art}=\text{max}(\Delta/{N}, 1{\text{Re}})$, where $\Delta$ represents the local mesh element size and $N$  is a tuning parameter that controls the strength of the dissipative effect. This formulation ensures that in high-resolution regions, the artificial viscosity remains negligible, preserving the physical accuracy of the simulation. These boundary conditions ensure minimal acoustic reflections while preserving the reliability of the numerical approach.

     \begin{table}  
        \centering
        \begin{tabular}{c c c c c c c c} 
                   & $N_{E}$ & $L_1$& $L_2$ & $l_s$ & $\alpha$ & $\sigma$ & $\omega$ \\ [0.9ex]
                M9 & 135885 & 3 & 4 & 3 & $4.5 \cdot 10^{-3}$ & 0.0349 & 2.6110 \\  
                M8 & 65465 & 3 & 4 & 3 & $4.5 \cdot 10^{-3 }$& 0.0347 & 2.6114 \\  
                M7 & 42521 & 4 & 4 & 3 & $4.5 \cdot 10^{-3}$ & 0.0352 & 2.6115 \\  
                M6 & 42594 & 10 & 10 & 8 & $3.2 \cdot 10^{-2}$ & 0.0342 & 2.6087 \\  
                M5 & 42573 & 2 & 3 & 2 & $2.0 \cdot 10^{-3 }$& 0.0333 & 2.6078 \\  
                M4 & 42188 & 3 & 4 & 2 & $4.0 \cdot 10^{-4}$ & 0.0348 & 2.6135 \\  
                M3 & 42188 & 3 & 4 & 2 & $2.0 \cdot 10^{-2}$ & 0.0347 & 2.6150 \\  
                M2 & 31129 & 3 & 4 & 3 & $4.5 \cdot 10^{-3 }$& 0.0378 & 2.6130 \\  
                M1 & 23476 & 3 & 4 & 3 & $4.5 \cdot 10^{-3}$ & 0.0415 & 2.6161 \\  
        
        \end{tabular}
        \caption{Meshes used in this work where $N_E$ is the number of cells, the normalized lengths $L_1$, $L_2$ and $l_s$ are indicated in figure \ref{Domain}, $\alpha$ is a constant defining the damping intensity and $\sigma$ and $\omega$ the normalized growth rate and frequency of the dominant mode.}
        \label{tablemesh}
        \end{table}

\section{Mean flow LNSE including all first order non-linear terms related to coherent fluctuations.}
\label{AC}
\begin{equation}
 \grad(\overline{\varrho}\overline{\boldsymbol{u}})=-2A^2\text{Re}\{\boldsymbol{\widetilde{u}}^*\cdot\grad\widetilde{\varrho} + \widetilde{\varrho}^*\grad\cdot\boldsymbol{\widetilde{u}}\}
\end{equation}
\begin{equation}
\boldsymbol{N}(\overline{\varrho},\overline{\boldsymbol{u}})=-2A^2\text{Re}\{\widetilde{\rho}^*i\lambda_i\boldsymbol{\widetilde{u}} + \widetilde{\rho}^*\boldsymbol{\widetilde{u}}\cdot\grad\boldsymbol{\overline{u}} + \widetilde{\rho}^*\boldsymbol{\overline{u}}\cdot\grad\boldsymbol{\widetilde{u}} + \overline{\rho}\boldsymbol{\widetilde{u}}^*\cdot\grad\boldsymbol{\widetilde{u}}\}
\end{equation}
$$ \hspace{-1.5cm} \boldsymbol{E}(\overline{\varrho},\overline{\boldsymbol{u}}, \overline{T}) = -2A^2\text{Re}\{\overline{\rho}\boldsymbol{\widetilde{u}}^*\cdot\grad\boldsymbol{\widetilde{T}} + \widetilde{\rho}^*\boldsymbol{\overline{u}}\cdot\grad\widetilde{T} + \widetilde{\rho}^*\boldsymbol{\widetilde{u}}\cdot\grad\overline{T}\  $$
$$ \hspace{1.7cm} +(\gamma-1)[\overline{\varrho}\widetilde{T}^*\grad\cdot\boldsymbol{\widetilde{u}} + \widetilde{\varrho}^*\widetilde{T}\grad\cdot\boldsymbol{\overline{u}} + \widetilde{\varrho}^*\overline{T}\grad\cdot\boldsymbol{\widetilde{u}}] $$
\begin{equation}
\hspace{1cm}
+ 
    \widetilde{\varrho}^*i\lambda_i\widetilde{T} - \gamma(\gamma-1) \frac{M^2}{\text{Re}}[\boldsymbol{\tau}(\widetilde{\boldsymbol{u}}^*):\boldsymbol{d}(\widetilde{\boldsymbol{u}})]
 \}
\end{equation}
\begin{equation}
    \overline{\varrho} \overline{T} -1 -\overline{P} = -2A^2\text{Re}\{ \widetilde{\varrho}^*\widetilde{T} \}
\end{equation}

\end{appendix}
\newpage
\bibliographystyle{jfm}
\bibliography{References}

\begin{thebibliography}{30}
\expandafter\ifx\csname natexlab\endcsname\relax\def\natexlab#1{#1}\fi
\def\au#1{#1} \def\ed#1{#1} \def\yr#1{#1}\def\at#1{#1}\def\jt#1{\textit{#1}}
  \def\bt#1{#1}\def\bvol#1{\textbf{#1}} \def\vol#1{#1} \def\pg#1{#1}
  \def\publ#1{#1}\def\arxiv#1{#1}\def\org#1{#1}\def\st#1{\textit{#1}}

\bibitem[Abdelmwgoud {\em et~al.\/}(2021)Abdelmwgoud, Shaaban \&
  Mohany]{10.1063/5.0051226}
{\sc \au{Abdelmwgoud, Moamenbellah}, \au{Shaaban, Mahmoud} \& \au{Mohany,
  Atef}} \yr{2021}  \at{Shear layer synchronization of aerodynamically isolated
  opposite cavities due to acoustic resonance excitation}.  \jt{Physics of
  Fluids}  \bvol{33}~(5),  \pg{055112}.

\bibitem[Abom(2010)]{book}
{\sc \au{Abom, Mats}} \yr{2010} {\em {An Introduction to Flow Acoustics}\/}.

\bibitem[{Ansys Inc.}(2024)]{ansys_fluent_2024}
{\sc \au{{Ansys Inc.}}} \yr{2024} {\em {Ansys\textsuperscript{\textregistered}
  Academic Research Fluent, Release 2024}\/}. {Ansys Inc.}, Canonsburg, PA,
  USA.

\bibitem[Boujo {\em et~al.\/}(2018)Boujo, Bauerheim \&
  Noiray]{saturation_boujo_2018}
{\sc \au{Boujo, Edouard}, \au{Bauerheim, Michaël} \& \au{Noiray, Nicolas}}
  \yr{2018}  \at{Saturation of a turbulent mixing layer over a cavity: response
  to harmonic forcing around mean flows}.  \jt{Journal of Fluid Mechanics}
  \bvol{853},  \pg{386–418}.

\bibitem[Bourquard {\em et~al.\/}(2021)Bourquard, Faure-Beaulieu \&
  Noiray]{Bourquard_Faure-Beaulieu_Noiray_2021}
{\sc \au{Bourquard, Claire}, \au{Faure-Beaulieu, Abel} \& \au{Noiray, Nicolas}}
  \yr{2021}  \at{Whistling of deep cavities subject to turbulent grazing flow:
  intermittently unstable aeroacoustic feedback}.  \jt{Journal of Fluid
  Mechanics}  \bvol{909},  \pg{A19}.

\bibitem[Citro {\em et~al.\/}(2015)Citro, Giannetti, Brandt \&
  Luchini]{Citro_Giannetti_Brandt_Luchini_2015}
{\sc \au{Citro, Vincenzo}, \au{Giannetti, Flavio}, \au{Brandt, Luca} \&
  \au{Luchini, Paolo}} \yr{2015}  \at{Linear three-dimensional global and
  asymptotic stability analysis of incompressible open cavity flow}.
  \jt{Journal of Fluid Mechanics}  \bvol{768},  \pg{113–140}.

\bibitem[Colonius(2004)]{Colonious}
{\sc \au{Colonius, Tim}} \yr{2004}  \at{{Modeling Artificial Boundary
  Conditions for Compressible Flow}}.  \jt{Annual Review of Fluid Mechanics}
  \bvol{36}.

\bibitem[Dai {\em et~al.\/}(2015)Dai, Jing \& Sun]{article}
{\sc \au{Dai, Xiwen}, \au{Jing, Xiaodong} \& \au{Sun, Xiaofeng}} \yr{2015}
  \at{Flow-excited acoustic resonance of a helmholtz resonator: Discrete vortex
  model compared to experiments}.  \jt{Physics of Fluids}  \bvol{27},
  \pg{057102}.

\bibitem[Fani {\em et~al.\/}(2018)Fani, Citro, Giannetti \&
  Auteri]{AcousticCyl}
{\sc \au{Fani, Andrea}, \au{Citro, Vincenzo}, \au{Giannetti, Flavio} \&
  \au{Auteri, F.}} \yr{2018}  \at{Computation of the bluff-body sound
  generation by a self-consistent mean flow formulation}.  \jt{Physics of
  Fluids}  \bvol{30},  \pg{036102}.

\bibitem[Faure-Beaulieu {\em et~al.\/}(2023)Faure-Beaulieu, Xiong, Pedergnana
  \& Noiray]{Faure-Beaulieu_Xiong_Pedergnana_Noiray_2023}
{\sc \au{Faure-Beaulieu, Abel}, \au{Xiong, Yuan}, \au{Pedergnana, Tiemo} \&
  \au{Noiray, Nicolas}} \yr{2023}  \at{Self-sustained azimuthal aeroacoustic
  modes. part 1. symmetry breaking of the mean flow by spinning waves}.
  \jt{Journal of Fluid Mechanics}  \bvol{971},  \pg{A21}.

\bibitem[Hanna \& Mohany(2023)]{HANNA2023103949}
{\sc \au{Hanna, Marc} \& \au{Mohany, Atef}} \yr{2023}  \at{Aeroacoustics and
  shear layer characteristics of confined cavities subject to low mach number
  flow}.  \jt{Journal of Fluids and Structures}  \bvol{121},  \pg{103949}.

\bibitem[Hecht(2012)]{FreeFEM}
{\sc \au{Hecht, F.}} \yr{2012}  \at{New development in freefem++}.  \jt{J.
  Numer. Math.}  \bvol{20}~(3-4),  \pg{251--265}.

\bibitem[Ho \& Kim(2021)]{Ho_Kim_2021}
{\sc \au{Ho, You~Wei} \& \au{Kim, Jae~Wook}} \yr{2021}  \at{A wall-resolved
  large-eddy simulation of deep cavity flow in acoustic resonance}.
  \jt{Journal of Fluid Mechanics}  \bvol{917},  \pg{A17}.

\bibitem[Hu {\em et~al.\/}(2008)Hu, Li \& Lin]{HU20084398}
{\sc \au{Hu, Fang~Q.}, \au{Li, X.D.} \& \au{Lin, D.K.}} \yr{2008}
  \at{Absorbing boundary conditions for nonlinear euler and navier–stokes
  equations based on the perfectly matched layer technique}.  \jt{Journal of
  Computational Physics}  \bvol{227}~(9),  \pg{4398--4424}.

\bibitem[Hussain \& Reynolds(1970)]{Hussain_Reynolds_1970}
{\sc \au{Hussain, A. K. M.~F.} \& \au{Reynolds, W.~C.}} \yr{1970}  \at{The
  mechanics of an organized wave in turbulent shear flow}.  \jt{Journal of
  Fluid Mechanics}  \bvol{41}~(2),  \pg{241–258}.

\bibitem[Koh {\em et~al.\/}(2003)Koh, Cho \& Moon]{koh2003aeroacoustic}
{\sc \au{Koh, Seung~Ryul}, \au{Cho, Yong} \& \au{Moon, Young~J}} \yr{2003}
  \at{Aeroacoustic computation of cavity flow in self-sustained oscillations}.
  \jt{KSME International Journal}  \bvol{17}~(4),  \pg{590--598}.

\bibitem[Li \& Yang(2025)]{Li_Yang_2025}
{\sc \au{Li, Zhaobin} \& \au{Yang, Xiaolei}} \yr{2025}  \at{Self-consistent
  model for active control of wind turbine wakes}.  \jt{Journal of Fluid
  Mechanics}  \bvol{1013},  \pg{A36}.

\bibitem[Mantic-Lugo \& Gallaire(2016)]{mantic2016self}
{\sc \au{Mantic-Lugo, Vladislav} \& \au{Gallaire, François}} \yr{2016}
  \at{Self-consistent model for the saturation mechanism of the response to
  harmonic forcing in the backward-facing step flow}.  \jt{Journal of Fluid
  Mechanics}  \bvol{793},  \pg{R1}.

\bibitem[Mantič-Lugo {\em et~al.\/}(2014)Mantič-Lugo, Arratia \&
  Gallaire]{PhysRevLett.113.084501}
{\sc \au{Mantič-Lugo, Vladislav}, \au{Arratia, Crist'obal} \& \au{Gallaire,
  François}} \yr{2014}  \at{Self-consistent mean flow description of the
  nonlinear saturation of the vortex shedding in the cylinder wake}.  \jt{Phys.
  Rev. Lett.}  \bvol{113},  \pg{084501}.

\bibitem[Meliga(2017)]{Meliga_2017}
{\sc \au{Meliga, Philippe}} \yr{2017}  \at{Harmonics generation and the
  mechanics of saturation in flow over an open cavity: a second-order
  self-consistent description}.  \jt{Journal of Fluid Mechanics}  \bvol{826},
  \pg{503–521}.

\bibitem[Meliga {\em et~al.\/}(2010)Meliga, Sipp \&
  Chomaz]{MELIGA_SIPP_CHOMAZ_2010}
{\sc \au{Meliga, Philippe}, \au{Sipp, Denis} \& \au{Chomaz, Jean-Marc}}
  \yr{2010}  \at{Effect of compressibility on the global stability of
  axisymmetric wake flows}.  \jt{Journal of Fluid Mechanics}  \bvol{660},
  \pg{499–526}.

\bibitem[Pirozzoli \& Colonius(2013)]{Colonious2013}
{\sc \au{Pirozzoli, Sergio} \& \au{Colonius, Tim}} \yr{2013}  \at{{Generalized
  characteristic relaxation boundary conditions for unsteady compressible flow
  simulations}}.  \jt{Journal of Computational Physics}  \bvol{248},
  \pg{109--126}.

\bibitem[Poinsot \& Lele(1992)]{POINSOT1992104}
{\sc \au{Poinsot, Thierry~J.} \& \au{Lele, Sanjiva~K.}} \yr{1992}  \at{Boundary
  conditions for direct simulations of compressible viscous flows}.
  \jt{Journal of Computational Physics}  \bvol{101}~(1),  \pg{104--129}.

\bibitem[Rockwell \& Naudascher(1978)]{Rockwell}
{\sc \au{Rockwell, Donald} \& \au{Naudascher, Eduard}} \yr{1978}
  \at{Review—self-sustaining oscillations of flow past cavities}.
  \jt{Journal of Fluids Engineering}  \bvol{100}~(2),  \pg{152--165}.

\bibitem[Rossiter(1964)]{rossiter1964wind}
{\sc \au{Rossiter, J.~E.}} \yr{1964}  \bt{Wind tunnel experiments on the flow
  over rectangular cavities at subsonic and transonic speeds}. {\em Tech.
  Rep.\/}.  \org{Royal Aircraft Establishment}.

\bibitem[Sierra-Ausin {\em et~al.\/}(2022)Sierra-Ausin, Fabre, Citro \&
  Giannetti]{Sierra-Ausin_Fabre_Citro_Giannetti_2022}
{\sc \au{Sierra-Ausin, Javier}, \au{Fabre, David}, \au{Citro, Vincenzo} \&
  \au{Giannetti, Flavio}} \yr{2022}  \at{Acoustic instability prediction of the
  flow through a circular aperture in a thick plate via an impedance
  criterion}.  \jt{Journal of Fluid Mechanics}  \bvol{943},  \pg{A48}.

\bibitem[Stuart(1958)]{Stuart_1958}
{\sc \au{Stuart, John~T.}} \yr{1958}  \at{On the non-linear mechanics of
  hydrodynamic stability}.  \jt{Journal of Fluid Mechanics}  \bvol{4}~(1),
  \pg{1–21}.

\bibitem[Wang {\em et~al.\/}(2024)Wang, Jia, He, He, Sung \&
  Liu]{Wang_Jia_He_He_Sung_Liu_2024}
{\sc \au{Wang, Peng}, \au{Jia, Sichang}, \au{He, Zheng}, \au{He, Chuangxin},
  \au{Sung, Hyung~Jin} \& \au{Liu, Yingzheng}} \yr{2024}  \at{Flow–acoustic
  resonance mechanism in tandem deep cavities coupled with acoustic eigenmodes
  in turbulent shear layers}.  \jt{Journal of Fluid Mechanics}  \bvol{984},
  \pg{A19}.

\bibitem[Yamouni {\em et~al.\/}(2013)Yamouni, Sipp \&
  Jacquin]{Yamouni_Sipp_Jacquin_2013}
{\sc \au{Yamouni, Sami}, \au{Sipp, Denis} \& \au{Jacquin, Laurent}} \yr{2013}
  \at{Interaction between feedback aeroacoustic and acoustic resonance
  mechanisms in a cavity flow: a global stability analysis}.  \jt{Journal of
  Fluid Mechanics}  \bvol{717},  \pg{134–165}.

\bibitem[Yim {\em et~al.\/}(2019)Yim, Meliga \& Gallaire]{yim2019self}
{\sc \au{Yim, Eunok}, \au{Meliga, Philippe} \& \au{Gallaire, Fran{\c{c}}ois}}
  \yr{2019}  \at{Self-consistent triple decomposition of the turbulent flow
  over a backward-facing step under finite amplitude harmonic forcing}.
  \jt{Proceedings of the Royal Society A: Mathematical, Physical and
  Engineering Sciences}  \bvol{475}~(2225),  \pg{20190018}.

\end{thebibliography}
\end{document}